\documentclass[onecolumn, a4size, 11pt]{IEEEtran}
\usepackage{amsmath}
\usepackage{amssymb}
\usepackage{amsfonts}
\usepackage{graphicx}
\usepackage{epsfig}
\usepackage{subfigure}
\usepackage{psfrag}

\title{Cooperative Multi-Cell Block Diagonalization with Per-Base-Station Power
Constraints\footnote{This paper has been presented in part at IEEE
Wireless Communications and Networking Conference (WCNC), Sydney,
Australia, April 18-21, 2010.}\footnote{R. Zhang is with the
Institute for Infocomm Research, A*STAR, Singapore
(e-mail:rzhang@i2r.a-star.edu.sg) and the Department of Electrical
and Computer Engineering, National University of Singapore
(e-mail:elezhang@nus.edu.sg).}}

\author{Rui Zhang}

\begin{document}
\maketitle \thispagestyle{empty}

\begin{abstract}
Block diagonalization (BD) is a practical linear precoding technique
that eliminates the inter-user interference in downlink multiuser
multiple-input multiple-output (MIMO) systems. In this paper, we
apply BD to the downlink transmission in a cooperative multi-cell
MIMO system, where the signals from different base stations (BSs) to
all the mobile stations (MSs) are jointly designed with the perfect
knowledge of the downlink channels and transmit messages.
Specifically, we study the optimal BD precoder design to maximize
the weighted sum-rate of all the MSs subject to a set of per-BS
power constraints. This design problem is formulated in an auxiliary
MIMO broadcast channel (BC) with a set of transmit power constraints
corresponding to those for individual BSs in the multi-cell system.
By applying convex optimization techniques, this paper develops an
efficient algorithm to solve this problem, and derives the
closed-form expression for the optimal BD precoding matrix. It is
revealed that the optimal BD precoding vectors for each MS in the
per-BS power constraint case are in general non-orthogonal, which
differs from the conventional orthogonal BD precoder design for the
MIMO-BC under one single sum-power constraint. Moreover, for the
special case of single-antenna BSs and MSs, the proposed solution
reduces to the optimal zero-forcing beamforming (ZF-BF) precoder
design for the weighted sum-rate maximization in the multiple-input
single-output (MISO) BC with  per-antenna power constraints.
Suboptimal and low-complexity BD/ZF-BF precoding schemes are also
presented, and their achievable rates are compared against those
with the optimal schemes.
\end{abstract}


\setlength{\baselineskip}{1.3\baselineskip}
\newtheorem{definition}{\underline{Definition}}[section]
\newtheorem{fact}{Fact}
\newtheorem{assumption}{Assumption}
\newtheorem{theorem}{\underline{Theorem}}[section]
\newtheorem{lemma}{\underline{Lemma}}[section]
\newtheorem{corollary}{\underline{Corollary}}[section]
\newtheorem{proposition}{\underline{Proposition}}[section]
\newtheorem{example}{\underline{Example}}[section]
\newtheorem{remark}{\underline{Remark}}[section]
\newtheorem{algorithm}{\underline{Algorithm}}[section]
\newcommand{\mv}[1]{\mbox{\boldmath{$ #1 $}}}

\section{Introduction}

The study of downlink beamforming and power control in cellular
systems has been an active area of research for many years.
Conventionally, most of the related works have focused on a
single-cell setup, where the co-channel interferences experienced by
the mobile stations (MSs) in a particular cell caused by the base
stations (BSs) of the other cells are treated as additional noises
at the receivers. For this setup, the downlink transmission in a
single cell with a multi-antenna BS and multiple
single-/multi-antenna MSs can be modeled by a multiple-input
single-/multiple-output (MISO/MIMO) broadcast channel (BC). It is
known that the dirty paper coding (DPC) technique achieves the
capacity region for the Gaussian MISO/MIMO BC, which constitutes all
the simultaneously achievable rates for all the MSs \cite{Shamai}.
However, DPC requires complicated nonlinear encoding and decoding
schemes and is thus difficult to implement in real-time systems.
Consequently, linear transmit and receive beamforming schemes for
the Gaussian MISO/MIMO BC have drawn a great deal of attention in
the literature \cite{Liu,Boche,Peel,Wiesel06,Hassibi}. In
particular, a simple linear precoding scheme for the MIMO BC is
known as block diagonalization (BD) \cite{Spencer04,Wong,Choi,Pan}.
With BD, the transmitted signal from the BS intended for each MS is
multiplied by a precoding matrix, which is restricted to be
orthogonal to the downlink channels associated with all the other
MSs. Thereby, all the inter-user interferences are eliminated and
each MS perceives an interference-free MIMO channel. In the special
case of MISO BC, BD reduces to the well-known zero-forcing
beamforming (ZF-BF) \cite{Peel}. Although BD is in general inferior
in terms of achievable rate as compared to the DPC-based optimal
nonlinear precoding scheme or the minimum-mean-squared-error
(MMSE)-based optimal linear precoding scheme, it performs very well
in the high signal-to-noise-ratio (SNR) regime and achieves the same
degrees of freedom (DoF) for the MISO-/MIMO-BC sum-rate as the
optimal linear/nonlinear precoding schemes \cite{Carie03}. Moreover,
BD can be generalized to incorporate nonlinear DPC processing, which
leads to a precoding scheme known as ZF-DPC \cite{Carie03}.

Recently, there has been a rapidly growing interest in shifting the
design paradigm from the conventional single-cell downlink
transmission to the multi-cell cooperative downlink transmission
\cite{Shamai01,Dai,Foschini,Somekh,Jing,Kaviani08}. In these
studies, it is assumed that BSs in a cellular network are connected
via backhaul links to a central processing unit (e.g., a dedicated
control station or a preassigned BS), which has the global knowledge
of transmit messages for all the MSs in the network and downlink
channels from each BS to all the MSs. Thereby, the central
processing unit is able to jointly design the downlink transmissions
for all BSs and provide them appropriate signals to transmit. As
demonstrated in these works, by utilizing the co-channel
interference across different cells in a coherent fashion, the
cooperative multi-cell downlink processing leads to enormous
throughput gains as compared to the conventional single-cell
processing with the co-channel interference treated as noise.
Moreover, design of distributed multi-cell downlink beamforming via
the use of belief propagation and message passing among BSs has also
been recently proposed in \cite{Hanly}, without the need of a
central controller.

In this work, we focus our study on the BD-based downlink precoding
for a fully cooperative multi-cell system equipped with a central
processing unit, which is assumed to have the perfect knowledge of
all downlink channels and transmit messages in the network. For this
setup, the BD precoding design problem can be formulated in an
auxiliary MISO/MIMO BC with the number of transmitting antennas
equal to the sum of those from all the cooperative BSs. However,
instead of adopting the conventional sum-power constraint for the
auxiliary MISO/MIMO BC as in prior works
\cite{Spencer04,Wong,Choi,Pan}, this paper applies a set of transmit
power constraints equivalent to those for individual BSs in the
multi-cell system. The BD precoder design problem subject to per-BS
power constraints is relatively new, and has been studied in, e.g.,
\cite{Huang,Hanzo09,Heath09}. In these works, the BD precoders are
designed essentially following the same principle as for the
conventional sum-power constraint case, i.e., the precoding vectors
known for the sum-power constraint case are adopted, and then power
allocation is optimization to maximize the sum-rate under per-BS
power constraints. However, it remains unclear whether the developed
BD precoder solutions therein are indeed optimal for the weighted
sum-rate maximization in a cooperative multi-cell system. In this
paper, we show that the BD precoder designs following the heuristic
of {\it separating} the beamforming design and power allocation
optimization are indeed suboptimal for rate maximization, while the
optimal BD precoder solution requires a new {\it joint} optimization
approach, as will be proposed in this paper.

It is worth noting that the computation problem for the achievable
rate region of the Gaussian MISO/MIMO BC subject to per-antenna
power constraints has been studied in \cite{Yu}. This work has been
recently extended in \cite{ZhangLan09a,Huh} to deal with more
general linear transmit power constraints for the MISO/MIMO BC, with
the per-antenna power constraint as a special case. The results in
\cite{Yu,ZhangLan09a,Huh} can be directly applied for a cooperative
multi-cell system to handle the per-BS power constraints, if the
DPC-based optimal nonlinear precoder or the MMSE-based optimal
linear precoder is used. On the other hand, the ZF-BF precoding
design, as a simplified version of BD for the case of MISO BC, has
been studied in \cite{Huang,Huh,Wiesel08} subject to per-antenna
power constraints. In \cite{Huang}, the ZF-BF precoding matrix is
taken as the pseudo inverse of the MISO-BC channel matrix and
thereby decomposes the MISO BC into parallel interference-free
scalar sub-channels for different MSs. The power allocation over the
sub-channels is then optimized under per-antenna power constraints.
However, it was pointed out in \cite{Wiesel08} that although the
ZF-BF precoding matrix for the MISO BC based on the channel pseudo
inverse is optimal for the sum-power constraint case, it is in
general suboptimal for the per-antenna power constraint case. Thus,
in \cite{Wiesel08} the authors proposed to apply the principle of
generalized matrix inverse to design the ZF-BF precoding with
per-antenna power constraints. The scheme in \cite{Wiesel08} has
been improved in terms of computational efficiency and extended to
the case of general linear power constraints in \cite{Huh}. However,
these MISO-BC ZF-BF solutions cannot be applied to obtain the
optimal BD precoder design for the more general MIMO BC with per-BS
power constraints.

The main contributions of this paper are summarized as follows:
\begin{itemize}
\item We formulate the MIMO-BC transmit optimization problem with
the BD precoding and equivalent per-BS power constraints as a convex
optimization problem. By applying convex optimization techniques, we
design an efficient algorithm to solve this problem. We also derive
the closed-form expression of the optimal BD precoding matrix to
maximize the weighted sum-rate for the MIMO-BC, from which we obtain
a lower bound on the number of BSs that should transmit with their
maximum power levels. More importantly, we prove that the optimal BD
precoding (beamforming) vectors for each MS in the case of per-BS
power constraints are in general {\it non-orthogonal}, which differs
from the conventional orthogonal BD precoder design for the
sum-power constraint case. Consequently, the orthogonal BD precoder
designs proposed in prior works \cite{Huang,Hanzo09,Heath09} for the
per-BS power constraint case are in general suboptimal (for weighted
sum-rate maximization).

\item For the special case of single-antenna BSs and MSs, we show
that the proposed BD precoding design for the MIMO-BC provides the
optimal ZF-BF precoder solution to maximize the weighted sum-rate
for the MISO BC with per-antenna power constraints. We also compare
the proposed solution with existing ones in prior works
\cite{Huh,Wiesel08} based on approaches such as the generalized
channel matrix inverse and the semi-definite programming (SDP) with
rank-one relaxation.

\item We also present a low-complexity, suboptimal scheme for the studied problem, which is obtained by computing the
conventional BD precoder design for the sum-power constraint case
with an optimal power allocation to meet the per-BS power
constraints. This scheme can be considered as an extension of that
given in \cite{Huang} for the MISO BC with the ZF-BF precoding and
per-antenna power constraints to the MIMO BC with the BD precoding
and per-BS power constraints. We derive an upper bound on the
maximum number of BSs transmitting with their full power levels for
this scheme, and identify the conditions under which this scheme
becomes sum-rate optimal.
\end{itemize}

The rest of this paper is organized as follows. Section
\ref{sec:system model} introduces the signal model for the downlink
transmission in a cooperative multi-cell system, and presents the
problem formulation for the weighted sum-rate maximization with the
BD precoding and per-BS power constraints. Section \ref{sec:optimal}
derives the optimal solution for this problem, and characterizes the
optimal solution for the special case of MISO BC with per-antenna
power constraints. Section \ref{sec:suboptimal} develops a heuristic
suboptimal scheme for the studied problem. Section
\ref{sec:simulation} provides numerical examples on the performance
of the proposed optimal and suboptimal schemes. Finally, Section
\ref{sec:conclusion} concludes the paper.

{\it Notations}: Scalars are denoted by lower-case letters, vectors
denoted by bold-face lower-case letters, and matrices denoted by
bold-face upper-case letters. $\mv{I}$ and $\mv{0}$  denote an
identity matrix and an all-zero matrix, respectively, with
appropriate dimensions. For a square matrix $\mv{S}$, ${\rm
Tr}(\mv{S})$, $|\mv{S}|$, $\mv{S}^{-1}$, and $\mv{S}^{1/2}$ denote
the trace, determinant, inverse (if $\mv{S}$ is full-rank), and
square-root of $\mv{S}$, respectively; and
$\mv{S}\succeq\mv{0}~(\mv{S}\preceq\mv{0})$ means that $\mv{S}$ is
positive (negative) semi-definite. ${\rm Diag}(\mv{a})$ denotes a
diagonal matrix with the main diagonal given by $\mv{a}$. For a
matrix $\mv{M}$ of arbitrary size, $\mv{M}^{H}$, $\mv{M}^{T}$, ${\rm
Rank}(\mv{M})$, and $\mv{M}^{\dag}$ denote the conjugate transpose,
transpose, rank, and pseudo inverse of $\mv{M}$, respectively.
$\mathbb{E}[\cdot]$ denotes the statistical expectation. The
distribution of a circularly symmetric complex Gaussian (CSCG)
random vector with mean vector $\mv{x}$ and covariance matrix
$\mv{\Sigma}$ is denoted by $\mathcal{CN}(\mv{x},\mv{\Sigma})$; and
$\sim$ stands for ``distributed as''. $\mathbb{C}^{x \times y}$
denotes the space of $x\times y$ complex matrices. $\|\mv{x}\|$
denotes the Euclidean norm of a complex vector $\mv{x}$.

\section{System Model and Problem Formulation}\label{sec:system model}

We consider a multi-cell system consisting of $A$ cells, each of
which has a BS to coordinate the transmission with $K_a$ MSs,
$K_a\geq 1$ and $a=1,\cdots,A$. Denote the total number of MSs in
the system as $K=\sum_{a=1}^AK_a$. For convenience, we assume that
all the BSs are equipped with the same number of antennas, denoted
by $M_B\geq 1$. Denote the total number of antennas across all the
BSs as $M=M_BA$. We also assume that each of $K$ MSs is equipped
with $N$ antennas, $N\geq 1$. Since we are interested in a fully
cooperative multi-cell system, the jointly designed downlink
transmission for all the BSs can be conveniently modeled by an
auxiliary MIMO BC with $M$ transmitting antennas and $K$ MSs each
having $N$ receiving antennas. For convenience, we assign the
indices to the transmitting antennas in the auxiliary MIMO BC
belonging to different BSs in the multi-cell system according to the
BS index, i.e., the $((a-1)M_B+1)$-th to $(aM_B)$-th antennas are
taken as the $M_B$ antennas from the $a$th BS, $a=1,\cdots,A$.
Similarly, the indices of MSs in the auxiliary MIMO BC are assigned
according to their cell indices, i.e., the
$(\sum_{i=1}^{a-1}K_i+1)$-th to $(\sum_{i=1}^{a}K_i)$-th MSs are
taken as the $K_a$ MSs from the $a$th cell, $a=1,\cdots,A$.
Accordingly, the discrete-time baseband signal of the auxiliary MIMO
BC is given by
\begin{align}\label{eq:signal model}
\mv{y}_{k}=\mv{H}_{k}\mv{x}_k+\sum_{j\neq k}\mv{H}_{k}\mv{x}_j+
\mv{z}_k, ~~ k=1,\cdots,K
\end{align}
where $\mv{x}_k\in\mathbb{C}^{M\times 1}$ and
$\mv{y}_k\in\mathbb{C}^{N\times 1}$ denote the transmitted and
received signals for the $k$th MS, respectively;
$\mv{H}_{k}\in\mathbb{C}^{N\times M}$ denotes the downlink channel
from all the $M$ base-station antennas to the $k$th MS; and
$\mv{z}_k\in\mathbb{C}^{N\times 1}$ denotes the receiver noise at
the $k$th MS. For convenience, we assume that
$\mv{z}_k\sim\mathcal{CN}(\mv{0},\mv{I}), \forall k$.

Without loss of generality, we can further express $\mv{x}_k$ as
\begin{equation}
\mv{x}_k=\mv{T}_k\mv{s}_k, ~~k=1,\ldots,K
\end{equation}
where $\mv{T}_k\in\mathbb{C}^{M\times D_k}$ is the precoding matrix
(which specifies both the transmit beamforming vectors and allocated
power values for different beams) for the $k$th MS; $D_k$ denotes
the number of transmitted data steams for the $k$th MS due to
spatial multiplexing, with $D_k\leq \min(M,N), \forall k$; and
$\mv{s}_k\in\mathbb{C}^{D_k\times 1}$ denotes the
information-bearing signal for the $k$th MS. We assume that
$\mv{s}_k$'s are independent over $k$. It is further assumed that a
Gaussian codebook is used for each MS at the transmitter and thus
$\mv{s}_k\sim\mathcal{CN}(\mv{0},\mv{I}), \forall k$. Denote
$\mv{S}_k=\mathbb{E}[\mv{x}_k\mv{x}_k^H]$ as the transmit covariance
matrix for the $k$th MS, with $\mv{S}_k\in\mathbb{C}^{M\times M}$
and $\mv{S}_k\succeq \mv{0}$. It is easy to verify that
$\mv{S}_k=\mv{T}_k\mv{T}_k^H$. The overall downlink transmit
covariance matrix for the $M$ cooperative transmitting antennas is
then $\mv{S}=\sum_{k=1}^K\mv{S}_k$. Since these transmitting
antennas come from more than one BS, they need to satisfy a set of
per-BS power constraints expressed as
\begin{align}\label{eq:per BS power constraint}
{\rm Tr}\left({\mv B}_a\mv{S}\right)\leq P ~ {\rm or}~
\sum_{k=1}^K{\rm Tr}\left({\mv B}_a\mv{S}_k\right)\leq P, ~~
a=1,\cdots,A
\end{align}
where
\begin{align}
\mv{B}_a\triangleq{\rm
Diag}\bigg(\underbrace{0,\cdots,0}_{(a-1)M_B},\underbrace{1,\cdots,1}_{M_B},\underbrace{0,\cdots,0}_{(A-a)M_B}\bigg)
\end{align}
and $P$ denotes the per-BS power constraint, which is assumed
identical for all the BSs. Note that in the special case of
single-antenna BSs and MSs, i.e., $M_B=N=1$, the per-BS power
constraints in (\ref{eq:per BS power constraint}) reduce to the
per-antenna power constraints for an equivalent MISO BC.

We assume a quasi-static fading environment and thus the channels of
interest in the auxiliary MIMO-BC remain constant for each downlink
transmission frame. We consider the BD precoding scheme for each
downlink frame transmission in the MIMO BC, which eliminates the
inter-user interference, i.e., in (\ref{eq:signal model}) we have
that for each given $k$, $\mv{H}_{j}\mv{x}_k=\mv{0}$ or
$\mv{H}_{j}\mv{T}_k=\mv{0}, \forall j\neq k$. It is easy to show
that the above ``ZF constraints'' are equivalent to the following
constraints
\begin{align}\label{eq:ZF constraints}
\mv{H}_j\mv{S}_k\mv{H}^H_j=\mv{0}, ~~ \forall j\neq k.
\end{align}
Assuming that the row vectors in all $\mv{H}_k$'s are linearly
independent (due to independent fading), from the constraints in
(\ref{eq:ZF constraints}) with $k=1,\ldots,K$, it follows that
$NK\leq M$ needs to be true in order to have a set of feasible
$\mv{S}_k$'s with $D_k={\rm Rank}(\mv{S}_k)=N, \forall k$, i.e., all
the MSs have the same number of data streams equal to $N$. In
practice, the total number of MSs in the system can be very large
such that the above condition is not satisfied. In such scenarios,
the transmissions to MSs can be scheduled into different time-slots
or frequency-bands by the central processing unit, where in each
time-slot/frequency-band, the number of MSs scheduled for
transmission satisfies the above condition. The interested readers
may refer to \cite{Yoo,Shen06} for the detailed design of downlink
transmission scheduling in the MISO/MIMO BC with ZF-BF/BD precoding.
For the rest of this paper, we assume that $D_k=N, \forall k$ and
$NK\leq M$ for simplicity.

We are now ready to present the weighted sum-rate maximization
problem for the downlink transmission in a cooperative multi-cell
system with the BD precoding and per-BS power constraints as
follows.
\begin{align}
\mathrm{(P1)}:~\mathop{\mathtt{max.}}_{\mv{S}_1,\cdots,\mv{S}_K} &
~~ \sum_{k=1}^Kw_k
\log\left|\mv{I}+\mv{H}_k\mv{S}_k\mv{H}_k^H\right|
\nonumber \\
\mathtt{s.t.} & ~~ \mv{H}_j\mv{S}_k\mv{H}^H_j=0, ~
\forall j\neq k \nonumber \\
&~~ \sum_{k=1}^K{\rm Tr}\left({\mv B}_a\mv{S}_k\right)\leq P,
~\forall a \nonumber \\ &~~ \mv{S}_k\succeq \mv{0}, ~ \forall k
\nonumber
\end{align}
where $w_k$ is the given non-negative rate weight for the $k$th MS.
For the purpose of exposition, we assume that $w_k>0, \forall k$.
Note that in (P1), we have used transmit covariance matrices
$\mv{S}_k$'s instead of precoding matrices $\mv{T}_k$'s as design
variables. This is because with $\mv{S}_k$'s, it is easy to verify
that (P1) is a convex optimization problem, since the objective
function is concave over $\mv{S}_k$'s and all the constraints
specify a convex set over $\mv{S}_k$'s. Thus, (P1) can be solved
using standard convex optimization techniques, e.g., the
interior-point method \cite{Boydbook}. However, such an approach
does not reveal the structure of the optimal BD precoding solution.
Therefore, in this paper we take a different approach to solve (P1),
which is based on the Lagrange duality method \cite{Boydbook} for
convex optimization problems. As will be shown in Section
\ref{sec:optimal}, this approach leads to a closed-form expression
for the optimal BD precoding matrix, and reveals some interesting
properties of the optimal solution.

\begin{remark}
It is worth noting that (P1) can be modified to incorporate
additional per-antenna power constraints at all the BSs. Let $P^{\rm
(pa)}$ denote the given per-antenna power threshold. Then, a set of
$M$ per-antenna power constraints can be included in (P1) as
follows:
\begin{align}
\sum_{k=1}^K{\rm Tr}\left(\mv{B}^{\rm (pa)}_i\mv{S}_k\right)\leq
P^{\rm (pa)}, ~~i=1,\cdots,M
\end{align}
where $\mv{B}^{\rm (pa)}_i$ is a diagonal matrix with the $i$th
diagonal element equal to one and all the others equal to zero.
Since the resulting optimization problem has similar structure to
(P1), it can be solved in a similar way. In this paper, we omit the
details for solving this modified version of (P1) for brevity.
\end{remark}

\begin{remark}
It is also worth noting that (P1) can be modified to solve the
weighted sum-rate maximization problem for the cooperative
multi-cell downlink transmission with the ZF-DPC precoding
\cite{Carie03} subject to the new per-BS/per-antenna power
constraints. With ZF-DPC, given a fixed encoding order for the
transmitted signals to different MSs (without loss of generality, we
assume that the encoding order is given by the MS index), the signal
for a later encoded MS is designed with the non-causal knowledge of
all the earlier encoded MS signals, of which the associated
interferences can be precanceled by DPC. By extending the ZF-DPC
scheme in \cite{Carie03} for the MISO BC to the case of MIMO BC,
(P1) can be modified to obtain the optimal ZF-DPC precoder design
subject to per-BS power constraints by rewriting the set of ZF
constraints in (P1) as
\begin{align}
\mv{H}_j\mv{S}_k\mv{H}^H_j=0, ~~ \forall j>k.
\end{align}
The resulting problem has similar structure to (P1) and can be
solved similarly (the details are omitted for brevity).
\end{remark}

\section{Proposed Solution} \label{sec:optimal}

In this section, we first present a new algorithm to solve (P1),
which reveals the structure of the optimal BD precoding matrix for
the general case with arbitrary numbers of antennas at the BS or MS.
Then, we investigate the developed solution for the special case of
single-antenna BSs and MSs, and compare it with other existing
solutions in the literature.

\subsection{General Case}

To solve (P1), it is desirable to remove the set of ZF constraints,
as follows: Define $\mv{G}_k=[\mv{H}_1^T,\cdots,\mv{H}_{k-1}^T,$
$\mv{H}_{k+1}^T,\cdots,\mv{H}_k^T]^T$, $k=1,\cdots,K$, where
$\mv{G}_k\in\mathbb{C}^{L\times M}$ with $L=N(K-1)$. Let the
(reduced) singular value decomposition (SVD) of $\mv{G}_k$ be
denoted as $\mv{G}_k=\mv{U}_k\mv{\Sigma}_k\mv{V}_k^H$, where
$\mv{U}_k\in\mathbb{C}^{L\times L}$ with
$\mv{U}_k^H\mv{U}_k=\mv{U}_k\mv{U}_k^H=\mv{I}$,
$\mv{V}_k\in\mathbb{C}^{M\times L}$ with
$\mv{V}_k^H\mv{V}_k=\mv{I}$, and $\mv{\Sigma}_k$ is a $L\times L$
positive diagonal matrix. Note that ${\rm Rank}(\mv{G}_k)=L<M$ under
the previous assumption that $NK\leq M$. Define the projection
matrix $\mv{P}_k=(\mv{I}-\mv{V}_k\mv{V}_k^H)$. Without loss of
generality, we can express
$\mv{P}_k=\tilde{\mv{V}}_k\tilde{\mv{V}}_k^H$, where
$\tilde{\mv{V}}_k\in\mathbb{C}^{M\times (M-L)}$ satisfies
$\mv{V}_k^H\tilde{\mv{V}}_k=\mv{0}$ and
$\tilde{\mv{V}}_k^H\tilde{\mv{V}}_k=\mv{I}$. Note that $[\mv{V}_k,
\tilde{\mv{V}}_k]$ forms a $M\times M$ unitary matrix. Then, we have
the following lemma.
\begin{lemma}\label{lemma:1}
The optimal solution of (P1) is given by
\begin{align}
\mv{S}_k=\tilde{\mv{V}}_k\mv{Q}_k\tilde{\mv{V}}_k^H, ~ k=1,\cdots,K
\end{align}
where $\mv{Q}_k\in\mathbb{C}^{(M-L)\times (M-L)}$ and
$\mv{Q}_k\succeq{\bf 0}$.
\end{lemma}
\begin{proof}
Please refer to Appendix \ref{appendix:proof 1}.
\end{proof}

\begin{remark}
In prior works \cite{Spencer04,Wong,Choi,Pan} on the design of BD
precoder for the MIMO BC with the sum-power constraint, it has been
observed that the columns (precoding vectors) in the BD precoding
matrix  for the $k$th MS, $\mv{T}_k$, with
$\mv{T}_k\mv{T}_k^H=\mv{S}_k$, should be linear combinations of
those in $\tilde{\mv{V}}_k$ in order to satisfy the constraints:
$\mv{H}_j\mv{T}_k=\mv{0}, \forall j\neq k$. Lemma \ref{lemma:1}
extends this result to the case of per-BS power constraints.
\end{remark}

With the optimal structures for $\mv{S}_k$'s given in Lemma
\ref{lemma:1}, it can be verified that all the ZF constraints in
(P1) are satisfied and thus can be removed. Thus, (P1) reduces to
the following equivalent problem
\begin{align}
\mathrm{(P2)}:~\mathop{\mathtt{max.}}_{\mv{Q}_1,\cdots,\mv{Q}_K} &
~~ \sum_{k=1}^Kw_k
\log\left|\mv{I}+\mv{H}_k\tilde{\mv{V}}_k\mv{Q}_k\tilde{\mv{V}}_k^H\mv{H}_k^H\right|
\nonumber \\
\mathtt{s.t.} &~~ \sum_{k=1}^K{\rm Tr}\left({\mv
B}_a\tilde{\mv{V}}_k\mv{Q}_k\tilde{\mv{V}}_k^H\right)\leq P,
~\forall a \nonumber \\ &~~ \mv{Q}_k\succeq \mv{0}, ~ \forall k
\nonumber.
\end{align}
Similar to (P1), it can be shown that (P2) is convex. Thus, (P2) is
solvable by the Lagrange duality method as follows. By introducing a
set of non-negative dual variables, $\mu_a, a=1,\cdots,A$,
associated with the set of per-BS power constraints in (P2), the
Lagrangian function of (P2) can be written as
\begin{align}\label{eq:Lagrangian}
L(\{\mv{Q}_k\},\{\mu_a\})=\sum_{k=1}^Kw_k
\log\left|\mv{I}+\mv{H}_k\tilde{\mv{V}}_k\mv{Q}_k\tilde{\mv{V}}_k^H\mv{H}_k^H\right|-\sum_{a=1}^A\mu_a\left(
\sum_{k=1}^K{\rm Tr}\left({\mv
B}_a\tilde{\mv{V}}_k\mv{Q}_k\tilde{\mv{V}}_k^H\right)-P\right)
\end{align}
where $\{\mv{Q}_k\}$ and $\{\mu_a\}$ denote the set of $\mv{Q}_k$'s
and the set of $\mu_a$'s, respectively. The Lagrange dual function
for (P2) is then defined as
\begin{align}\label{eq:dual}
g(\{\mu_a\})=\max_{\mv{Q}_k\succeq \mv{0}, \forall k}
L(\{\mv{Q}_k\},\{\mu_a\}).
\end{align}
Moreover, the dual problem of (P2) is defined as
\begin{align}
\mathrm{(P2-D)}:~\min_{\mu_a\geq 0, \forall a} ~g(\{\mu_a\}).
\nonumber
\end{align}
Since (P2) is convex and satisfies the Slater's condition
\cite{Boydbook}, the duality gap between the optimal objective value
of (P2) and that of (P2-D) is zero. Thus, (P2) can be solved
equivalently by solving (P2-D). Moreover, (P2-D) is convex and can
be solved by the subgradient-based method, e.g., the ellipsoid
method \cite{BGT81}, given the fact that the subgradient of function
$g(\{\mu_a\})$ at a set of fixed $\mu_a$'s is $P-\sum_{k=1}^K{\rm
Tr}\left({\mv
B}_a\tilde{\mv{V}}_k\mv{Q}_k^{\star}\tilde{\mv{V}}_k^H\right)$ for
$\mu_a$, $a=1,\cdots,A$, where $\{\mv{Q}_k^{\star}\}$ is the optimal
solution for the maximization problem in (\ref{eq:dual}) with the
given set of $\mu_a$'s.

Next, we focus on solving for $\{\mv{Q}_k^{\star}\}$ with a set of
fixed $\mu_a$'s. From (\ref{eq:Lagrangian}), it is observed that the
maximization problem in (\ref{eq:dual}) can be separated into $K$
independent subproblems each involving only one $\mv{Q}_k$. By
discarding the irrelevant terms, the corresponding subproblem, for a
given $k$, can be expressed as
\begin{align}
\mathrm{(P3)}:~\max_{\mv{Q}_k\succeq \mv{0}} w_k
\log\left|\mv{I}+\mv{H}_k\tilde{\mv{V}}_k\mv{Q}_k\tilde{\mv{V}}_k^H\mv{H}_k^H\right|-
{\rm Tr}\left({\mv
B}_{\mu}\tilde{\mv{V}}_k\mv{Q}_k\tilde{\mv{V}}_k^H\right) \nonumber
\end{align}
where ${\mv B}_{\mu}\triangleq\sum_{a=1}^A\mu_a\mv{B}_a$. Note that
${\mv B}_{\mu}$ is a diagonal matrix with the diagonal elements
given by different $\mu_a$'s in the order of $a=1,\cdots,A$. We then
have the following lemma.
\begin{lemma}\label{lemma:2}
Let $A_{\mu}$ denote the number of $\mu_a$'s in the main diagonal of
$\mv{B}_{\mu}$, $a\in\{1,\cdots,A\}$, with $\mu_a>0$. Then, for (P3)
to have a bounded objective value, it holds that $A_{\mu}\geq
\lceil\frac{M-N(K-1)}{M_B}\rceil$.
\end{lemma}
\begin{proof}
Please refer to Appendix \ref{appendix:proof 2}.
\end{proof}

\begin{remark}\label{remark:lower bound}
It is noted that by applying the Karash-Kuhn-Tucker (KKT) conditions
\cite{Boydbook} to (P2), the fact that $\mu_a>0$ for a given
$a\in\{1,\cdots,A\}$ implies that the corresponding power constraint
must be tight with the optimal solution for $\{\mv{Q}_k\}$.
Accordingly, in (P1) the optimal downlink transmit covariance
matrices $\mv{S}_k$'s must make the $a$th per-BS power constraint
tight. Therefore, Lemma \ref{lemma:2} provides a lower bound on the
number of BSs for which the corresponding transmit power constraints
must be tight with the optimal $\mv{S}_k$'s for (P1).
\end{remark}

With Lemma \ref{lemma:2} and $L=N(K-1)$, we can assume without loss
of generality that $M_BA_{\mu}\geq (M-L)$ since we are only
interested in the case where the objective value of (P3) and that of
(P1) are both bounded. Accordingly, we have ${\rm
Rank}(\tilde{\mv{V}}_k^H\mv{B}_{\mu}\tilde{\mv{V}}_k)=\min(M_BA_{\mu},M-L)=
M-L$. Thus,
$\tilde{\mv{V}}_k^H\mv{B}_{\mu}\tilde{\mv{V}}_k\in\mathbb{C}^{(M-L)\times(M-L)}$
is a full-rank matrix and its inverse exists. Moreover, since ${\rm
Tr}(\mv{XY})={\rm Tr}(\mv{YX})$, in (P3) we have ${\rm Tr}({\mv
B}_{\mu}\tilde{\mv{V}}_k\mv{Q}_k\tilde{\mv{V}}_k^H)={\rm
Tr}((\tilde{\mv{V}}_k^H\mv{B}_{\mu}\tilde{\mv{V}}_k)^{1/2}\mv{Q}_k(\tilde{\mv{V}}_k^H\mv{B}_{\mu}\tilde{\mv{V}}_k)^{1/2})$.
We thus define
\begin{align}
\tilde{\mv{Q}}_k=(\tilde{\mv{V}}_k^H\mv{B}_{\mu}\tilde{\mv{V}}_k)^{1/2}\mv{Q}_k(\tilde{\mv{V}}_k^H\mv{B}_{\mu}\tilde{\mv{V}}_k)^{1/2}.
\end{align}
Then, (P3) can be reformulated to maximize
\begin{align}\label{eq:P3 new}
w_k
\log\left|\mv{I}+\mv{H}_k\tilde{\mv{V}}_k(\tilde{\mv{V}}_k^H\mv{B}_{\mu}\tilde{\mv{V}}_k)^{-1/2}\tilde{\mv{Q}}_k
(\tilde{\mv{V}}_k^H\mv{B}_{\mu}\tilde{\mv{V}}_k)^{-1/2}
\tilde{\mv{V}}_k^H\mv{H}_k^H\right|-{\rm
Tr}\left(\tilde{\mv{Q}}_k\right)
\end{align}
subject to $\tilde{\mv{Q}}_k\succeq \mv{0}$. Note that ${\rm
Rank}(\mv{H}_k\tilde{\mv{V}}_k(\tilde{\mv{V}}_k^H\mv{B}_{\mu}\tilde{\mv{V}}_k)^{-1/2})=\min(N,M-L)=
N$. Thus, the following (reduced) SVD can be obtained as
\begin{align}\label{eq:SVD}
\mv{H}_k\tilde{\mv{V}}_k(\tilde{\mv{V}}_k^H\mv{B}_{\mu}\tilde{\mv{V}}_k)^{-1/2}=\hat{\mv{U}}_k\hat{\mv{\Sigma}}_k\hat{\mv{V}}_k^H
\end{align}
where $\hat{\mv{U}}_k\in\mathbb{C}^{N\times N}$,
$\hat{\mv{V}}_k\in\mathbb{C}^{(M-L)\times N}$, and
$\hat{\mv{\Sigma}}_k={\rm
Diag}(\hat{\sigma}_{k,1},\cdots,\hat{\sigma}_{k,N})$. Substituting
the above SVD into (\ref{eq:P3 new}) and applying the Hadamard's
inequality (see, e.g., \cite{Cover}) yields the following optimal
solution for (\ref{eq:P3 new}) as
$\tilde{\mv{Q}}_k^{\star}=\hat{\mv{V}}_k\mv{\Lambda}_k\hat{\mv{V}}_k^H$,
where $\mv{\Lambda}_k={\rm
Diag}(\lambda_{k,1},\cdots,\lambda_{k,N})$, where $\lambda_{k,i}$,
$i=1,\cdots,N$, can be obtained by the standard water-filling
algorithm \cite{Cover} as
\begin{align}\label{eq:optimal power}
\lambda_{k,i}=\left(w_k-\frac{1}{\hat{\sigma}_{k,i}^2}\right)^+
\end{align}
where $(x)^+\triangleq\max(0,x)$. To summarize, the optimal solution
of (P3) for a given set of $\mu_a$'s can be expressed as
\begin{align}\label{eq:optimal Qk}
\mv{Q}_k^{\star}=(\tilde{\mv{V}}_k^H\mv{B}_{\mu}\tilde{\mv{V}}_k)^{-1/2}
\hat{\mv{V}}_k\mv{\Lambda}_k\hat{\mv{V}}_k^H(\tilde{\mv{V}}_k^H\mv{B}_{\mu}\tilde{\mv{V}}_k)^{-1/2},
~ k=1,\cdots,K.
\end{align}
Note that when the optimal solution for $\{\mu_a\}$ in (P2-D) is
obtained, the corresponding solution in (\ref{eq:optimal Qk})
becomes optimal for (P2). By combining this result with Lemma
\ref{lemma:1}, we obtain the following theorem.
\begin{theorem}\label{theorem:1}
The optimal solution of (P1) is given by
\begin{align}\label{eq:optimal Sk}
\mv{S}_k^{\star}=\tilde{\mv{V}}_k(\tilde{\mv{V}}_k^H\mv{B}_{\mu}^{\star}\tilde{\mv{V}}_k)^{-1/2}
\hat{\mv{V}}_k\mv{\Lambda}_k\hat{\mv{V}}_k^H(\tilde{\mv{V}}_k^H\mv{B}_{\mu}^{\star}\tilde{\mv{V}}_k)^{-1/2}\tilde{\mv{V}}_k^H,
~ k=1,\cdots,K
\end{align}
where $\mv{B}_{\mu}^{\star}=\sum_{a=1}^A\mu_a^{\star}\mv{B}_a$, with
$\mu_a^{\star}$'s being the optimal dual solutions of (P2).
\end{theorem}

The algorithm for solving (P1) is summarized as follows.

\underline{Algorithm (A1)}:
\begin{itemize}
\item {\bf Initialize} $\mu_a\geq0, a=1,\cdots,A$.
\item {\bf Repeat}
\begin{itemize}
\item[1.] Solve $\mv{Q}_k^{\star}, k=1,\cdots,K$ using
(\ref{eq:optimal Qk}) with the given $\mu_a$'s;
\item[2.] Compute the subgradient of $g(\{\mu_a\})$ as $P-\sum_{k=1}^K{\rm Tr}\left({\mv
B}_a\tilde{\mv{V}}_k\mv{Q}_k^{\star}\tilde{\mv{V}}_k^H\right),
a=1,\cdots,A$, and update $\mu_a$'s accordingly based on the
ellipsoid method \cite{BGT81};
\end{itemize}
\item {\bf Until} all the $\mu_a$'s converge to a prescribed accuracy.
\item {\bf Set}
$\mv{S}_k^{\star}=\tilde{\mv{V}}_k\mv{Q}_k^{\star}\tilde{\mv{V}}_k^H,
k=1,\cdots,K$.
\end{itemize}

From Theorem \ref{theorem:1} and the fact that
$\mv{S}_k=\mv{T}_k\mv{T}_k^H, \forall k$, we obtain the following
corollary.
\begin{corollary}
The optimal BD precoding matrices to maximize the weighted sum-rate
for the MIMO-BC subject to the per-BS power constraints in
(\ref{eq:per BS power constraint}) are given by
\begin{align}\label{eq:optimal Tk}
\mv{T}_k^{\star}=\tilde{\mv{V}}_k(\tilde{\mv{V}}_k^H\mv{B}_{\mu}^{\star}\tilde{\mv{V}}_k)^{-1/2}
\hat{\mv{V}}_k\mv{\Lambda}_k^{1/2}, ~k=1,\ldots,K.
\end{align}
\end{corollary}

In the following remarks, we discuss some interesting observations
on the optimal BD precoding matrices given by (\ref{eq:optimal Tk}).

\begin{remark}[{\it Channel Diagonalization}] One desirable property of linear
precoding for a point-to-point MIMO channel is that the precoding
matrix, when jointly deployed with a unitary decoding matrix at the
receiver, is able to diagonalize the MIMO channel into parallel
scalar sub-channels, over which independent encoding and decoding
can be applied to simplify the transceiver design. Here, we verify
that the optimal $\mv{T}_k^{\star}$ given in (\ref{eq:optimal Tk})
satisfies this ``channel diagonalization'' property, as follows:
\begin{align}
\mv{H}_k\mv{T}_k^{\star}&=\mv{H}_k\tilde{\mv{V}}_k(\tilde{\mv{V}}_k^H\mv{B}_{\mu}^{\star}\tilde{\mv{V}}_k)^{-1/2}
\hat{\mv{V}}_k\mv{\Lambda}_k^{1/2} \\
&=
\hat{\mv{U}}_k\hat{\mv{\Sigma}}_k\hat{\mv{V}}_k^H\hat{\mv{V}}_k\mv{\Lambda}_k^{1/2}
\label{eq:line a}
\\ &=\hat{\mv{U}}_k\hat{\mv{\Sigma}}_k\mv{\Lambda}_k^{1/2}
\end{align}
where (\ref{eq:line a}) is due to (\ref{eq:SVD}). Therefore, with a
unitary decoding matrix $\hat{\mv{U}}_k^H$ applied at the $k$th MS
receiver, the MIMO channel for the $k$th MS with BD precoding is
diagonalized into $N$ scalar sub-channels with channel gains given
by the main diagonal of the diagonal matrix
$\hat{\mv{\Sigma}}_k\mv{\Lambda}_k^{1/2}$. It is easy to verify that
the above linear precoder and decoder processing preserves the
single-user MIMO channel capacity.
\end{remark}

\begin{remark}[{\it Comparison with Conventional Sum-Power Constraint}] It is
noted that (P1) can be modified to deal with the case where a single
sum-power constraint over all the BSs (instead of a set of per-BS
power constraints) is applied. This can be done via replacing the
set of per-BS power constraints in (P1) by
\begin{align}\label{eq:sum power constraint}
\sum_{k=1}^K{\rm Tr}(\mv{S}_k)\leq P^{\rm (sum)}
\end{align}
where $P^{\rm (sum)}$ denotes the given sum-power constraint. Note
that (P1) in this case corresponds to the conventional BD precoder
design problem for the MIMO BC with a sum-power constraint as
studied in  \cite{Spencer04,Wong,Choi,Pan}. It can be shown that the
developed solution for (P1) can be applied to this case, while the
corresponding matrix $\mv{B}_{\mu}^{\star}$ should be modified as
$\mu^{\star}\mv{I}$ with $\mu^{\star}$ denoting the optimal dual
solution associated with the sum-power constraint in (\ref{eq:sum
power constraint}). From (\ref{eq:optimal Sk}), it follows that the
optimal solution for this modified version of (P1) is given by
\begin{align}\label{eq:optimal Sk new}
\mv{S}_k^{\star\star}=\frac{1}{\mu^{\star}}\tilde{\mv{V}}_k
\hat{\mv{V}}_k\mv{\Lambda}_k\hat{\mv{V}}_k^H\tilde{\mv{V}}_k^H, ~
k=1,\cdots,K.
\end{align}
Moreover, from (\ref{eq:SVD}) with
$\mv{B}_{\mu}^{\star}=\mu^{\star}\mv{I}$, it follows that
$\hat{\mv{V}}_k$ is obtained from the SVD:
$\frac{1}{\sqrt{\mu^{\star}}}\mv{H}_k\tilde{\mv{V}}_k=\hat{\mv{U}}_k\hat{\mv{\Sigma}}_k\hat{\mv{V}}_k^H$
and is thus independent of $\mu^{\star}$. Accordingly, the optimal
precoding matrix in the sum-power constraint case is
$\mv{T}_k^{\star\star}=\frac{1}{\sqrt{\mu^{\star}}}\tilde{\mv{V}}_k
\hat{\mv{V}}_k\mv{\Lambda}_k^{1/2}$. Comparing
$\mv{T}_k^{\star\star}$ with  $\mv{T}_k^{\star}$ in (\ref{eq:optimal
Tk}) for the per-BS power constraint case, we see that
$\mv{T}_k^{\star\star}$ consists of {\it orthogonal} columns
(beamforming vectors) since
$\hat{\mv{V}}_k^H\tilde{\mv{V}}_k^H\tilde{\mv{V}}_k
\hat{\mv{V}}_k=\mv{I}$, while $\mv{T}_k^{\star}$ in general consists
of {\it non-orthogonal} columns if $\mv{B}_{\mu}^{\star}$ is a
non-identity diagonal matrix (i.e., the optimal $\mu_a^{\star}$'s
are not all equal). This is the very reason that the BD precoder
designs in prior works \cite{Huang,Hanzo09,Heath09} based on the
orthogonal precoder structure $\mv{T}_k^{\star\star}$  are in
general suboptimal for the per-BS power constraint case.
\end{remark}

\subsection{Special Case: MISO BC with Per-Antenna Power Constraints}

In this subsection, we investigate the developed solution for the
special case of $M_B=N=1$, where the auxiliary MIMO BC with the
per-BS power constraints reduces to an equivalent MISO BC with the
corresponding per-antenna power constraints, and the BD precoding
reduces to the ZF-BF precoding. With $N=1$, $\mv{H}_k$ is a
row-vector, which we denote by $\mv{h}_k\in\mathbb{C}^{M\times 1},
k=1,\cdots,K$. Accordingly, the SVD in (\ref{eq:SVD}) is rewritten
as
\begin{align}\label{eq:SVD MISO}
\mv{h}_k^H\tilde{\mv{V}}_k(\tilde{\mv{V}}_k^H\mv{B}_{\mu}\tilde{\mv{V}}_k)^{-1/2}=\hat{\sigma}_k\hat{\mv{v}}_k^H
\end{align}
where $\hat{\sigma}_k>0$ and
$\hat{\mv{v}}_k\in\mathbb{C}^{(M-L)\times 1}$.  From
(\ref{eq:optimal power}), (\ref{eq:optimal Sk}), and (\ref{eq:SVD
MISO}), it follows that the optimal downlink transmit covariance
matrix for the $k$th MS, $\mv{S}_k^{\star}$, in the case of $N=1$ is
expressed as
\begin{align}
\mv{S}_k^{\star}&=~\lambda_k\tilde{\mv{V}}_k(\tilde{\mv{V}}_k^H\mv{B}_{\mu}^{\star}\tilde{\mv{V}}_k)^{-1/2}
\hat{\mv{v}}_k\hat{\mv{v}}_k^H(\tilde{\mv{V}}_k^H\mv{B}_{\mu}^{\star}\tilde{\mv{V}}_k)^{-1/2}\tilde{\mv{V}}_k^H
\label{eq:optimal Sk MISO 1} \\&=~
~\lambda_k\hat{\sigma}_k^{-2}\tilde{\mv{V}}_k(\tilde{\mv{V}}_k^H\mv{B}_{\mu}^{\star}\tilde{\mv{V}}_k)^{-1}\tilde{\mv{V}}_k^H\mv{h}_k\mv{h}_k^H
\tilde{\mv{V}}_k(\tilde{\mv{V}}_k^H\mv{B}_{\mu}^{\star}\tilde{\mv{V}}_k)^{-1}\tilde{\mv{V}}_k^H
\label{eq:optimal Sk MISO 2}
\end{align}
where $\lambda_k=(w_k-1/\hat{\sigma}_k^2)^+$. It is thus easy to
observe that in this case ${\rm Rank}(\mv{S}_k^{\star})\leq 1$.
Thus, the corresponding optimal precoding matrix reduces to a
(beamforming) vector denoted by
$\mv{t}_k^{\star}\in\mathbb{C}^{M\times 1}$, where
$\mv{S}_k^{\star}=\mv{t}_k^{\star}(\mv{t}_k^{\star})^H$ and
\begin{align}\label{eq:optimal tk}
\mv{t}_k^{\star}=\lambda_k^{1/2}\hat{\sigma}_k^{-1}\tilde{\mv{V}}_k(\tilde{\mv{V}}_k^H\mv{B}_{\mu}^{\star}\tilde{\mv{V}}_k)^{-1}\tilde{\mv{V}}_k^H\mv{h}_k.
\end{align}
Note that (\ref{eq:optimal tk}) holds regardless of $M_B$, but
$M_B=1$ corresponds to the per-antenna power constraint case for the
MISO BC. Furthermore, the optimal beamforming vector for the $k$th
MS in the conventional sum-power constraint case (with
$\mv{B}_{\mu}^{\star}=\mu^{\star}\mv{I}$) is obtained from
(\ref{eq:optimal tk}) as
\begin{align}\label{eq:optimal tk sum power}
\mv{t}_k^{\star\star}=\lambda_k^{1/2}\hat{\sigma}_k^{-1}(\mu^{\star})^{-1}\tilde{\mv{V}}_k\tilde{\mv{V}}_k^H
\mv{h}_k.
\end{align}

In the following, we discuss some interesting observations on the
optimal ZF-BF precoding design in (\ref{eq:optimal tk}), as compared
with other prior results reported in \cite{Peel,Huh,Wiesel08}.

\begin{remark}\label{remark:comparison}
Denote $\mv{T}=[\mv{t}_1,\cdots,\mv{t}_K]\in \mathbb{C}^{M\times K}$
as the precoding matrix for a MISO BC with $M$ transmitting antennas
and $K$ single-antenna receiving MSs. Then, for the sum-power
constraint case with $\mv{t}_k=\mv{t}_k^{\star\star}$ given in
(\ref{eq:optimal tk sum power}), the corresponding optimal precoding
matrix $\mv{T}^{\star\star}$ becomes the conventional ZF-BF design
for the MISO BC based on the channel pseudo inverse \cite{Peel},
i.e., $\mv{T}^{\star\star}$ can be put in the form
$\mv{T}^{\star\star}=\mv{H}^{\dag}\hat{\mv{\Lambda}}$, where
$\mv{H}=[\mv{h}_1,\cdots,\mv{h}_K]^H$ and $\hat{\mv{\Lambda}}={\rm
Diag}(\hat{\lambda}_1,\cdots,\hat{\lambda}_K)$, where
$\hat{\lambda}_k=\lambda_k^{1/2}\hat{\sigma}_k, k=1,\cdots,K$.
However, it is observed that the ZF-BF design based on the channel
pseudo inverse is in general suboptimal for the MISO BC with the
per-antenna/per-BS power constraints, where the optimal precoding
matrix $\mv{T}^{\star}$ is obtained with $\mv{t}_k=\mv{t}_k^{\star}$
given in (\ref{eq:optimal tk}). Note that $\mv{t}_k^{\star}$ becomes
collinear with $\mv{t}_k^{\star\star}$ regardless of
$\mu_a^{\star}$'s when $M=K$. In this case, $\tilde{\mv{V}}_k$
becomes a vector, $\tilde{\mv{v}}_k\in\mathbb{C}^{M\times 1}$, and
$\mv{t}_k^{\star}$ and $\mv{t}_k^{\star\star}$ can both be written
in the form $p_k\tilde{\mv{v}}_k$, with $p_k\geq 0$. Furthermore, it
can be shown that this result holds regardless of the value of $M_B$
provided that $N=1$ and $M=M_BA=K$.
\end{remark}

\begin{remark}
In \cite{Wiesel08}, the authors proposed a ZF-BF precoding design
for the MISO BC with per-antenna power constraints in the form of
the generalized inverse of $\mv{H}$. The corresponding precoding
matrix is expressed as
\begin{align}\label{eq:generalized inverse}
\mv{T}=[\mv{g}_1a_1,\cdots,\mv{g}_Ka_K]+\mv{U}^{\perp}[\mv{b}_1,\cdots,\mv{b}_K]
\end{align}
where $\mv{g}_k$ is the normalized (to unit-norm) $k$th column in
$\mv{H}^{\dag}$, $k=1,\cdots,K$;
$\mv{U}^{\perp}\in\mathbb{C}^{M\times(M-K)}$ is a projection matrix
onto the orthogonal complement of the space spanned by the row
vectors in $\mv{H}$, $(\mv{U}^{\perp})^H\mv{U}^{\perp}=\mv{I}$;
$a_k$'s and $\mv{b}_k$'s are design variables, $k=1,\cdots,K$. In
other words, each beamforming vector $\mv{t}_k$ in $\mv{T}$ given by
(\ref{eq:generalized inverse}) is a linear combination of $\mv{g}_k$
and the columns in $\mv{U}^{\perp}$. We see that the beamforming
vectors given in (\ref{eq:generalized inverse}) are in accordance
with the optimal $\mv{t}_k^{\star}$'s given in (\ref{eq:optimal tk})
due to the fact that for the MISO BC with $N=1$ and thus
$L=M-N(K-1)=M-K+1$, the space spanned by the columns in
$\tilde{\mv{V}}_k\in\mathbb{C}^{M\times L}$ is the same as that
spanned by $\mv{g}_k$ and the columns in $\mv{U}^{\perp}$. Note that
in \cite{Huh}, an algorithm is proposed to obtain the ZF-BF
precoding matrix for the MISO BC with per-antenna power constraints
by numerically searching over $a_k$'s and $\mv{b}_k$'s in
(\ref{eq:generalized inverse}). In this paper, the optimal ZF-BF
precoders are found based on the closed-form expression in
(\ref{eq:optimal tk}) and applying a numerical search over the dual
variable $\mu_a$'s by the ellipsoid method.
\end{remark}

\begin{remark}
It is also worth comparing the proposed method for solving (P1) in
the MISO BC case with the method presented in \cite{Wiesel08}. For
the method in \cite{Wiesel08}, a set of transmit beamforming
vectors, $\mv{t}_1,\cdots,\mv{t}_K$, are used in a MISO BC. Thus,
the weighted sum-rate maximization problem with the ZF-BF precoding
and per-antenna power constraints can be formulated as
\begin{align}
\mathrm{(P4)}:~ \mathop{\mathtt{max.}}_{\mv{t}_1,\cdots,\mv{t}_K} &
~~ \sum_{k=1}^Kw_k \log\left(1+\|\mv{h}_k^H\mv{t}_k\|^2\right)
\nonumber \\
\mathtt{s.t.} & ~~ \mv{h}_j^H\mv{t}_k=0, \forall j\neq k \nonumber \\
&~~ \sum_{k=1}^K{\rm Tr}\left({\mv B}_i\mv{t}_k\mv{t}_k^H\right)\leq
P, ~\forall i \nonumber
\end{align}
where ${\mv B}_i\in\mathbb{C}^{M\times M}$ denotes a diagonal matrix
with the $i$th diagonal element equal to one and all the others
equal to zero, $i=1,\cdots,M$; and $P$ refers to the per-antenna
power constraint. Note that (P4) is non-convex due to the fact that
the objective function is not necessarily concave over $\mv{t}_k$'s.
In \cite{Wiesel08}, it is proposed to convert (P4) into an
equivalent problem in terms of
$\mv{S}_k\triangleq\mv{t}_k\mv{t}_k^H, k=1,\cdots,K$, which is
expressed as
\begin{align}
\mathrm{(P5)}:~ \mathop{\mathtt{max.}}_{\mv{S}_1,\cdots,\mv{S}_K} &
~~ \sum_{k=1}^Kw_k \log\left(1+\mv{h}_k^H\mv{S}_k\mv{h}_k\right)
\nonumber \\
\mathtt{s.t.} & ~~ \mv{h}_j^H\mv{S}_k\mv{h}_j=0, ~
\forall j\neq k \nonumber \\
&~~ \sum_{k=1}^K{\rm Tr}\left({\mv B}_i\mv{S}_k\right)\leq P,
~\forall i \nonumber \\ &~~ \mv{S}_k\succeq \mv{0}, ~ \forall k
\nonumber \\ &~~ {\rm Rank}(\mv{S}_k)=1, ~\forall k. \nonumber
\end{align}
Note that (P5) can be treated as (P1) in the case of $N=1$ and
$M_B=1$ (thus $M=M_BA=A$), and with an additional set of rank-one
constraints for $\mv{S}_k$'s. However, these rank-one constraints
are non-convex and thus render (P5) non-convex in general. As a
special form of (P1), (P5) without the rank-one constraints is
convex, and thus can be solved efficiently by, e.g., the
interior-point method \cite{Boydbook}. However, the obtained
solution for $\mv{S}_k$ is not guarantied to be rank-one. In
\cite{Wiesel08}, it is proved that there always exists a solution
that consists of a set of rank-one $\mv{S}_k$'s for (P5), and a
method is provided to construct the rank-one solution from the
corresponding solution (with rank greater than one) of (P5) without
the rank-one constraints. In contrast, the proposed method in this
paper obtains the closed-form solution for (P5), which, as given in
(\ref{eq:optimal Sk MISO 2}), is guaranteed to be rank-one.
\end{remark}

\section{Suboptimal Solution} \label{sec:suboptimal}

In this section, we propose a suboptimal solution for (P1), which
can be obtained with less complexity than the optimal solution.
First, we define the projected channel of $\mv{H}_k$ associated with
the projection matrix $\mv{P}_k$ as
$\mv{H}_k^{\perp}=\mv{H}_k\mv{P}_k=\mv{H}_k\tilde{\mv{V}}_k\tilde{\mv{V}}_k^H,
k=1,\cdots,K$, where $\mv{H}_k^{\perp}\in\mathbb{C}^{N\times M}$,
and ${\rm Rank}(\mv{H}_k^{\perp})=\min(N,M-L)=N$. Next, define the
(reduced) SVD of $\mv{H}_k^{\perp}$ as
\begin{align}\label{eq:SVD suboptimal}
\mv{H}_k^{\perp}=\mv{U}_k^{\perp}\mv{\Sigma}_k^{\perp}(\mv{V}_k^{\perp})^H
\end{align}
where $\mv{U}_k^{\perp}\in\mathbb{C}^{N\times N}$,
$\mv{\Sigma}_k^{\perp}={\rm
Diag}(\sigma_{k,1}^{\perp},\cdots,\sigma_{k,N}^{\perp})$, and
$\mv{V}_k^{\perp}\in\mathbb{C}^{M\times N}$. Then, the proposed
suboptimal solution for (P1) is give by
\begin{align}\label{eq:suboptimal Sk}
\bar{\mv{S}}_k=\mv{V}_k^{\perp}\bar{\mv{\Lambda}}_k(\mv{V}_k^{\perp})^H
\end{align}
where $\bar{\mv{\Lambda}}_k={\rm
Diag}(\bar{\lambda}_{k,1},\cdots,\bar{\lambda}_{k,N})$ denotes the
power allocation for the $k$th MS. It is worth noting that the above
solution for $\mv{S}_k$ is in general suboptimal for (P1) by
comparing it with the optimal solution in (\ref{eq:optimal Sk}).
Also note that (\ref{eq:suboptimal Sk}) is optimal for the sum-power
constraint case as discussed in Section \ref{sec:optimal}-A, since
it can be shown that in (\ref{eq:optimal Sk new})
$\tilde{\mv{V}}_k\hat{\mv{V}}_k=\mv{V}_k^{\perp}$ with
$\mv{B}_{\mu}^{\star}=\mu^{\star}\mv{I}$. With $\bar{\mv{S}}_k$'s
given in (\ref{eq:suboptimal Sk}), it can be shown that the ZF
constraints in (P1) are satisfied and thus can be removed;
furthermore, in the objective function of (P1), the following
equalities hold:
\begin{align}
&\log\left|\mv{I}+\mv{H}_k\bar{\mv{S}}_k\mv{H}_k^H\right| \\ =&
\log\left|\mv{I}+(\mv{H}^{\perp}_k+\mv{H}_k\mv{V}_k\mv{V}_k^H)\bar{\mv{S}}_k(\mv{H}^{\perp}_k+\mv{H}_k\mv{V}_k\mv{V}_k^H)^H\right|
\label{eq:a}
\\ =&
\log\left|\mv{I}+\mv{H}_k^{\perp}\bar{\mv{S}}_k(\mv{H}_k^{\perp})^H\right|
\label{eq:b}
\\ =&
\log\left|\mv{I}+\mv{U}_k^{\perp}\mv{\Sigma}_k^{\perp}\bar{\mv{\Lambda}}_k\mv{\Sigma}_k^{\perp}(\mv{U}_k^{\perp})^H\right|
\label{eq:c}
\\ =&
\log\left|\mv{I}+(\mv{\Sigma}_k^{\perp})^2\bar{\mv{\Lambda}}_k\right|
\label{eq:d}
\end{align}
where $(\ref{eq:a})$ is due to the fact that
$\tilde{\mv{V}}_k\tilde{\mv{V}}_k^H+\mv{V}_k\mv{V}_k^H=\mv{I}$;
$(\ref{eq:b})$ is due to the fact that
$\bar{\mv{S}}_k\mv{V}_k=\mv{0}$ since
$\tilde{\mv{V}}_k^H\mv{V}_k=\mv{0}$;  $(\ref{eq:c})$ is due to
(\ref{eq:SVD suboptimal}) and (\ref{eq:suboptimal Sk}); and
$(\ref{eq:d})$ is due to the fact that
$\log|\mv{I}+\mv{XY}|=\log|\mv{I}+\mv{YX}|$. From (\ref{eq:d}), we
see that the MIMO channel for the $k$th MS is diagonalized into $N$
scalar sub-channels with channel gains given by
$\bar{\lambda}_{k,i}, i=1,\cdots,N$. Accordingly, (P1) is reduced to
the following problem
\begin{align}
\mathrm{(P6)}:~ \mathop{\mathtt{max.}}_{\{\bar{\lambda}_{k,i}\}} &
~~ \sum_{k=1}^Kw_k\sum_{i=1}^N
\log\left(1+(\sigma^{\perp}_{k,i})^2\bar{\lambda}_{k,i}\right)
\nonumber \\
\mathtt{s.t.} &~~ \sum_{k=1}^K\sum_{i=1}^N\|\mv{v}_k^{\perp}[a,i]\|^2\bar{\lambda}_{k,i}\leq P, ~\forall a \nonumber \\
&~~ \bar{\lambda}_{k,i}\geq 0, ~ \forall k,i \nonumber
\end{align}
where $\{\bar{\lambda}_{k,i}\}$ denotes the set of
$\bar{\lambda}_{k,i}$'s, $k=1,\cdots,K$ and $i=1,\cdots,N$, while
$\mv{v}_k^{\perp}[a,i]$ denotes the vector consisting of the
elements from the $i$th column and the $((a-1)M_B+1)$-th to
$(aM_B)$-th rows in $\mv{V}_k^{\perp}, a=1,\cdots,A$ and
$i=1,\cdots,N$. It can be verified that (P6) is a convex
optimization problem. Thus, similar to (P2), the Lagrange duality
method can be applied to solve (P6) by introducing a set of dual
variables, $\mu_a, a=1,\cdots,A$, associated with the set of per-BS
power constraints in (P6). For brevity, we omit here the details for
the derivation and present the optimal solution (power allocation)
for $\{\bar{\lambda}_{k,i}\}$ as follows:
\begin{align}\label{eq:suboptimal power}
\bar{\lambda}_{k,i}=\left(\frac{w_k}{\sum_{a=1}^A\mu_a\|\mv{v}_k^{\perp}[a,i]\|^2}-\frac{1}{(\sigma^{\perp}_{k,i})^2}\right)^+.
\end{align}
Similar to (A1), the following algorithm can be used to obtain the
proposed suboptimal solution for (P1).

\underline{Algorithm (A2)}:
\begin{itemize}
\item {\bf Initialize} $\mu_a\geq 0, a=1,\cdots,A$.
\item {\bf Compute} the SVDs:
$\mv{H}_k\tilde{\mv{V}}_k\tilde{\mv{V}}_k^H=\mv{U}_k^{\perp}\mv{\Sigma}_k^{\perp}(\mv{V}_k^{\perp})^H,
k=1,\cdots,K$.
\item {\bf Repeat}
\begin{itemize}
\item[1.] Solve $\{\bar{\lambda}_{k,i}\}$ using
(\ref{eq:suboptimal power}) with the given $\mu_a$'s;
\item[2.] Compute the subgradient of the dual function for (P6) as
$P-\sum_{k=1}^K\sum_{i=1}^N\|\mv{v}_k^{\perp}[a,i]\|^2\bar{\lambda}_{k,i}$,
$a=1,\cdots,A$, and update $\mu_a$'s accordingly based on the
ellipsoid method \cite{BGT81};
\end{itemize}
\item {\bf Until} all the $\mu_a$'s converge to a prescribed accuracy.
\item {\bf Set}
$\bar{\mv{S}}_k=\mv{V}_k^{\perp}\bar{\mv{\Lambda}}_k(\mv{V}_k^{\perp})^H,
k=1,\cdots,K$.
\end{itemize}

As compared with (A1), (A2) has a lower complexity due to the fact
that for each loop in the ``Repeat'', only the power allocation
computation in (\ref{eq:suboptimal power}) is implemented, instead
of the precoding matrix computation given in (\ref{eq:optimal Qk}).
Due to the suboptimal structure of the downlink transmit covariance
matrix in (\ref{eq:suboptimal Sk}) for (A2) as compared to the
optimal one in (\ref{eq:optimal Sk}) for (A1), (A2) in general leads
to a suboptimal solution for (P1), except in the special case of
$N=1$ and $M=K$ where the transmit covariance structure in
(\ref{eq:suboptimal Sk}) is known to be optimal (see Remark
\ref{remark:comparison}). In this special case, (A2) can be used as
an alternative algorithm to (A1) to obtain the optimal solution for
(P1).

\begin{remark}
It is worth noting that (A2) can be shown equivalent to the
algorithm proposed in \cite{Huang} for the special case of
$M_B=N=1$, i.e., the MISO BC with the ZF-BF precoding and the
per-antenna power constraints. In this case, similar to Remark
\ref{remark:comparison}, the proposed suboptimal solution in
(\ref{eq:suboptimal Sk}) corresponds to a precoding matrix in the
form $\mv{T}=\mv{H}^{\dag}\mv{\Theta}$, where $\mv{H}$ denotes the
downlink MISO-BC channel, and $\mv{\Theta}$ is a diagonal matrix
with the main diagonal that has been optimized in a similar way as
we have used for (P6). According to our previous discussions, this
algorithm is indeed suboptimal for (P1) if $M>K$.
\end{remark}

At last, as a counterpart of Lemma \ref{lemma:2}, we have the
following lemma.
\begin{lemma}\label{lemma:3}
Let $A^*$ denote the number of active per-BS power constraints with
the optimal solution for (P6). It then holds that $A^*\leq NK$.
\end{lemma}
\begin{proof}
Please refer to Appendix \ref{appendix:proof 3}.
\end{proof}
Lemma \ref{lemma:3} provides an upper bound on the number of active
per-BS power constraints for the suboptimal solution of (P1)
obtained by (A2). It follows that in the case of $(A/NK)\gg1$, most
of the BSs in the cooperative multi-cell system cannot transmit with
their full power levels with the suboptimal BD precoder design
obtained by (A2).

\section{Numerical Examples}\label{sec:simulation}

In this section, we provide numerical examples to illustrate the
results in this paper. For the purpose of exposition, we assume that
the channel $\mv{H}_k$'s in (\ref{eq:signal model}) are independent
over $k$, and all the elements in each channel matrix are
independent CSCG random variables with zero mean and unit variance.
Moreover, we consider the sum-rate maximization for the cooperative
multi-cell downlink transmission, i.e., $w_k$'s are all equal to one
in (P1). The obtained numerical results along with related
discussions are presented in the following subsections.

\subsection{Convergence Behavior}

\begin{figure}
\centering
 \epsfxsize=0.7\linewidth
    \includegraphics[width=12cm]{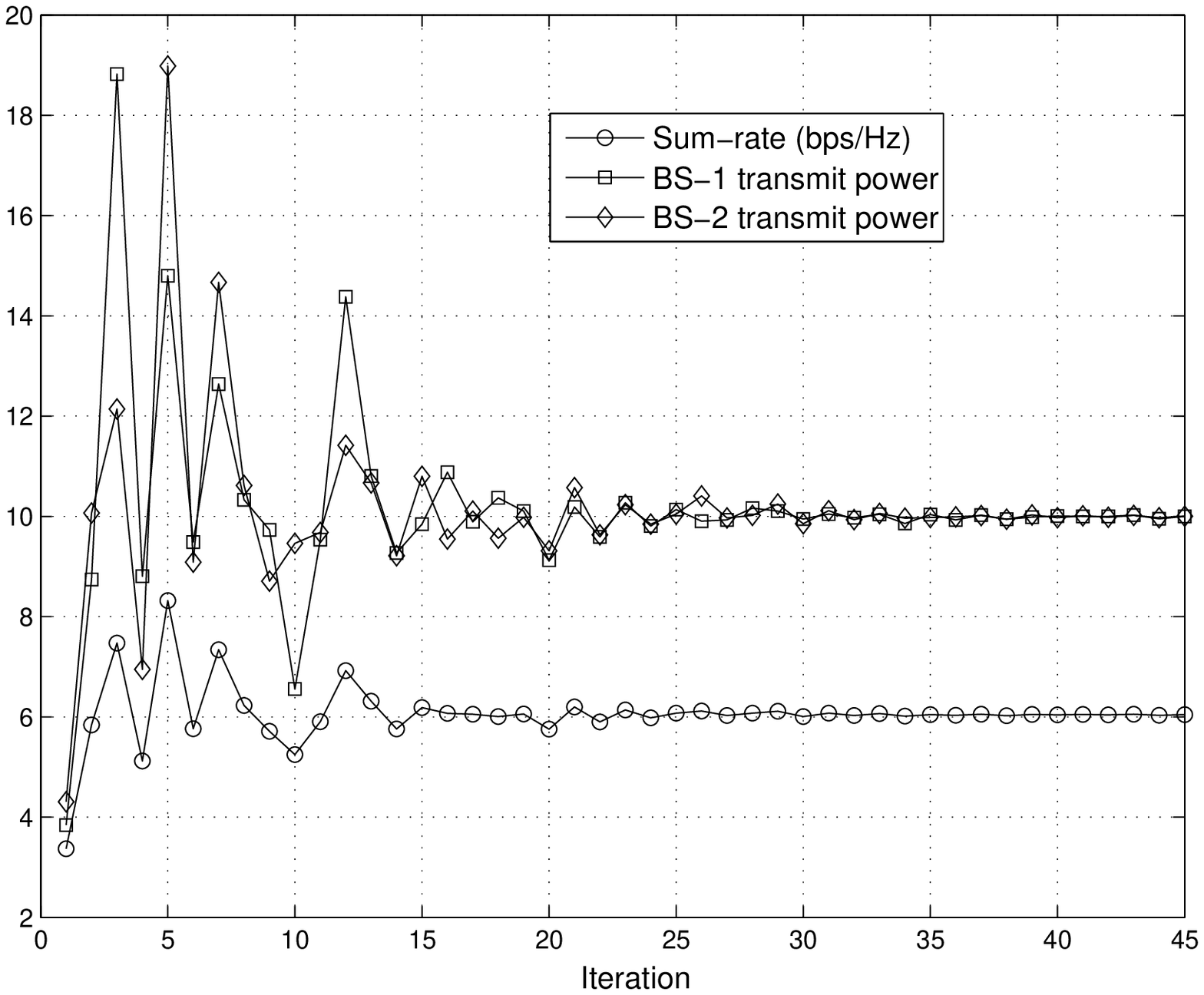}
\caption{Convergence behavior of Algorithm (A1).}
\label{fig:converge}
\end{figure}

In Fig. \ref{fig:converge}, we show the convergence behavior of
Algorithm (A1) for solving (P1). It is assumed that $A=2, M_B=4,
K=4$, and $N=2$. The transmit power constraint $P$ for each of the
two BSs is set equal to $10$. The initial values assigned to
$\mu_a$'s in (A1) are $\mu_1=\mu_2=0.2$. The achievable sum-rate and
the consumed transmit powers by the two BSs are shown for different
iterations in (A1), each with a pair of updated values for $\mu_1$
and $\mu_2$. As observed, the plotted rate and power values all
converge to fixed values after around $30$ iterations. The converged
transmit powers for the two BSs are observed both equal to their
given constraint value, which is $10$. A similar convergence
behavior for Algorithm (A2) can be observed and thus omitted here.
Generally speaking, the convergence speed of both (A1) and (A2)
depends critically on the total number of per-BS power constraints,
$A$, which is also the number of dual variable $\mu_a$'s to be
searched. With the ellipsoid method, it is known that the complexity
for searching $\mu_a$'s in (A1) or (A2) is $\mathcal{O}(A^2)$ for
large values of $A$ \cite{BGT81}. Thus, the number of iterations for
the algorithm convergence grows asymptotically in the order of the
square of the number of BSs in the system.

\subsection{MISO BC with Per-Antenna Power Constraints}

\begin{figure}
\centering
 \epsfxsize=0.7\linewidth
    \includegraphics[width=12cm]{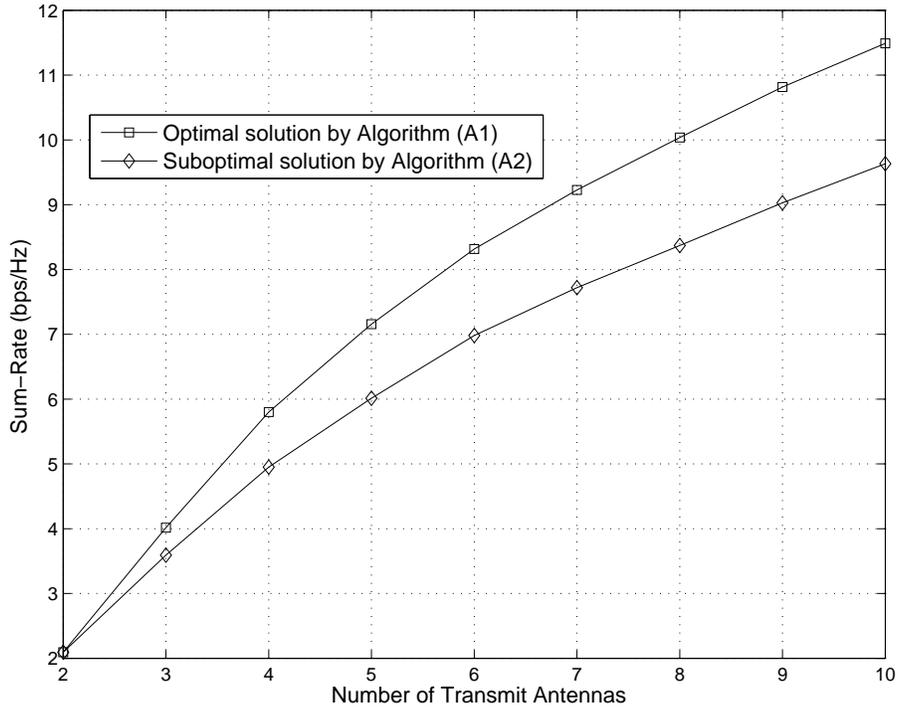}
\caption{Comparison of the sum-rate in the MISO BC with the ZF-BF
precoding and the per-antenna power constraints.} \label{fig:MISO}
\end{figure}

\begin{figure}
\centering
 \epsfxsize=0.7\linewidth
    \includegraphics[width=12cm]{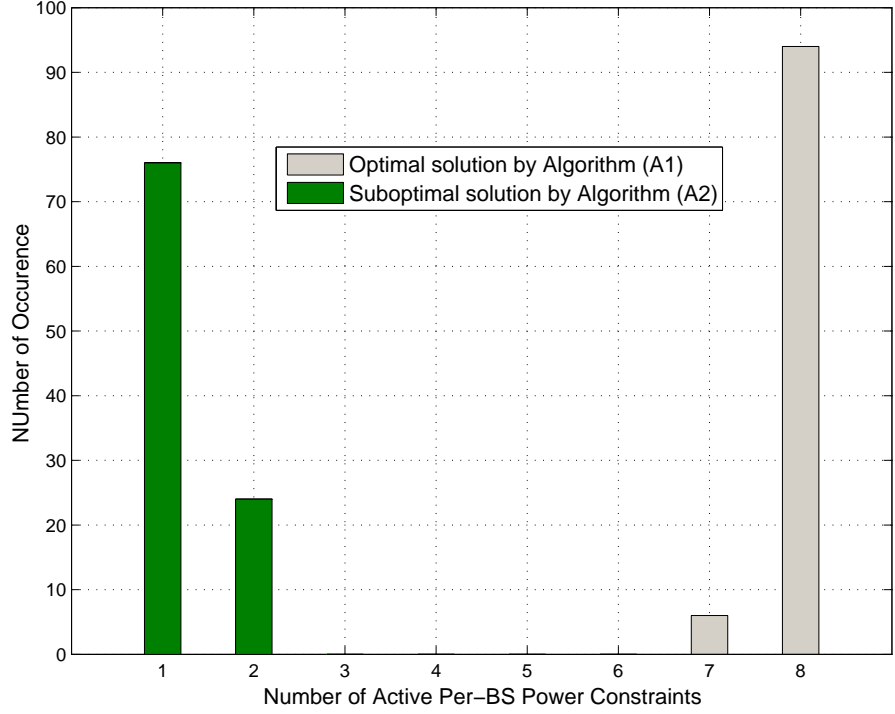}
\caption{Comparison of the number of active per-BS power constraints
in the MISO BC with the ZF-BF precoding and the per-antenna power
constraints.} \label{fig:count}
\end{figure}

Next, we consider a special case of the cooperative multi-cell
downlink transmission with $M_B=N=1$, which is equivalent to a MISO
BC with the corresponding per-antenna power constraints. The
per-BS/per-antenna power constraint is assumed to be $P=10$. In Fig.
\ref{fig:MISO}, we compare the achievable sum-rate with the optimal
ZF-BF precoder obtained by (A1) against that with the suboptimal
precoder obtained by (A2). The number of MSs is fixed as $K=2$,
while the total number of transmitting antennas $M$ ranges from $2$
to $10$. It is observed that when $M=K=2$, the achievable rates for
both the optimal and suboptimal precoders are identical, which is in
accordance with our discussions in Section \ref{sec:suboptimal}. It
is also observed that when $M>K$, the sum-rate gain of the optimal
precoder solution over the suboptimal solution increases with $M$.
In order to explain this observation, in Fig. \ref{fig:count} we
show the histograms for the number of active per-antenna power
constraints with the optimal and suboptimal solutions over $100$
random MISO-BC realizations for the case of $M=8$. It is observed
that the number of active per-antenna power constraints with the
optimal solution is always no less than
$\lceil\frac{M-N(K-1)}{M_B}\rceil=7$, while that with the suboptimal
solution is always no larger than $NK=2$, in accordance with Lemmas
\ref{lemma:2} and \ref{lemma:3}, respectively. We thus see that when
$M$ becomes much larger than $K$ for the MISO BC, the optimal ZF-BF
design can utilize the full transmit powers from at least $(M-K+1)$
antennas, while the suboptimal design can only have at most $K$
antennas transmitting with their full powers. This explains why in
Fig. \ref{fig:MISO} the rate gap between the optimal and suboptimal
ZF-BF designs enlarges as $M$ increases with a fixed $K$.

\subsection{MIMO BC with Per-Antenna Power Constraints}

\begin{figure}
\centering
 \epsfxsize=0.7\linewidth
    \includegraphics[width=12cm]{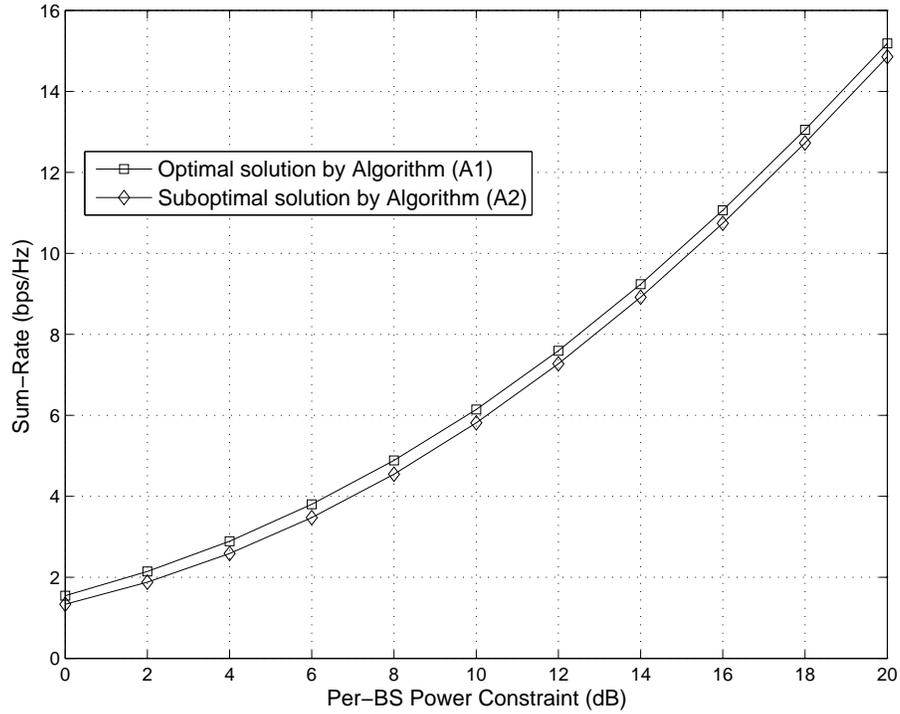}
\caption{Comparison of the sum-rate in the MIMO BC with the BD
precoding and the per-antenna/per-BS power constraints.}
\label{fig:MIMO}
\end{figure}

Last, we consider the case of multi-antenna MS receivers. For the
corresponding auxiliary MIMO BC, we assume that $A=4, M_B=1, K=2$,
and $N=2$. Note that in this case although $M=NK$, i.e., the total
number of transmitting antennas are equal to that of receiving
antennas, (A2) in general leads to a suboptimal solution for (P1)
due to the fact that $N>1$. In Fig. \ref{fig:MIMO}, we show the
achievable sum-rates for both the optimal and suboptimal BD
precoders vs. the per-BS/per-antenna transmit power constraint $P$.
It is observed that although the optimal precoder solution still
performs better than the suboptimal one, their rate gap is not as
large as that in Fig. \ref{fig:MISO} when $M>NK$ and $P=10$. This is
due to the fact that in the case of $M=NK$, although the maximum
number of antennas transmitting with full powers for the suboptimal
solution is still limited by $NK$ according to Lemma \ref{lemma:3},
such a constraint is not useful since $M_B=1$ and $A=NK$. The
practical rule of thumb here is that when $M_B=1$ and $A$ is not
substantially larger than $NK$, the low-complexity suboptimal BD
precoder obtained by (A2) can be applied to achieve the sum-rate
performance close to that of the optimal BD precoder obtained by
(A1).

\section{Conclusion} \label{sec:conclusion}

This paper studies the design of block diagonalization (BD) linear
precoder for the fully cooperative multi-cell downlink transmission
subject to individual power constraints for the base stations (BSs).
By applying convex optimization techniques, this paper derives the
closed-form expression for the optimal BD precoding matrix to
maximize the weighted sum-rate of all users in the multi-cell
system. The optimal BD precoding vectors for each user are shown to
be in general non-orthogonal, which differs from the conventional
orthogonal precoder design for the sum-power constraint case. A
suboptimal heuristic method is also proposed, which combines the
conventional orthogonal BD precoder design with an optimized power
allocation to meet the per-BS power constraints. Furthermore, this
paper shows that the proposed optimal BD precoder solution provides
the optimal zero-forcing beamforming (ZF-BF) solution for the
special case of MISO BC with per-antenna power constraints. The
results in this paper are readily extended to obtain the optimal BD
precoders for the MIMO-BC with general linear transmit power
constraints, which include per-antenna/per-BS power constraints as
special cases.

\appendices

\section{Proof of Lemma \ref{lemma:1}}\label{appendix:proof 1}

Let $\{\mv{S}_1^{\star},\cdots,\mv{S}_K^{\star}\}$ denote the
optimal solution of (P1). Without loss of generality, for any given
$k\in\{1,\cdots,K\}$, we can express $\mv{S}_k^{\star}$ in the
following form
\begin{align}
\mv{S}_k^{\star}&=[\tilde{\mv{V}}_k, \mv{V}_k]
\left[\begin{array}{ll} \mv{A} &\mv{B} \\ \mv{B}^H & \mv{C}
\end{array} \right] [\tilde{\mv{V}}_k,
\mv{V}_k]^H  \label{eq:line 1} \\ &=
\tilde{\mv{V}}_k\mv{A}\tilde{\mv{V}}_k^H+\tilde{\mv{V}}_k\mv{B}\mv{V}_k^H
+ \mv{V}_k\mv{B}^H\tilde{\mv{V}}_k^H+ \mv{V}_k\mv{C}\mv{V}_k^H
\label{eq:line 2}
\end{align}
where $\mv{A}\in\mathbb{C}^{(M-L)\times(M-L)}$,
$\mv{B}\in\mathbb{C}^{(M-L)\times L}$, and
$\mv{C}\in\mathbb{C}^{L\times L}$. Note that $\mv{A}=\mv{A}^H$ and
$\mv{C}=\mv{C}^H$. Since $\mv{S}_k^{\star}$ must satisfy the set of
ZF constraints in (P1), it follows that
\begin{align}
\mv{V}_k^H\mv{S}_k^{\star}\mv{V}_k=0. \label{eq:line 3}
\end{align}
From (\ref{eq:line 2}) and (\ref{eq:line 3}), it follows that
$\mv{C}=\mv{0}$. Furthermore, from the theory of Schur complement
\cite{Boydbook}, it is known that $\mv{S}_k^{\star}\succeq\mv{0}$ if
and only if  (iff) the following conditions are satisfied:
\begin{align}
\mv{A}&\succeq\mv{0}
\\ (\mv{I}-\mv{A}\mv{A}^{\dag})\mv{B}&=\mv{0} \label{eq:line 4} \\
\mv{C}-\mv{B}^H\mv{A}^{\dag}\mv{B}&\succeq\mv{0}. \label{eq:line 5}
\end{align}
Since $\mv{A}\succeq\mv{0}$, it follows that
$\mv{B}^H\mv{A}^{\dag}\mv{B}\succeq\mv{0}$. Using this fact together
with $\mv{C}=\mv{0}$, from (\ref{eq:line 5}) it follows that
$\mv{B}^H\mv{A}^{\dag}\mv{B}=\mv{0}$. Thus, from (\ref{eq:line 4})
it follows that $\mv{B}=\mv{0}$. With $\mv{B}=\mv{0}$ and
$\mv{C}=\mv{0}$, from (\ref{eq:line 2}) it follows that
$\mv{S}_k^{\star}=\tilde{\mv{V}}_k\mv{A}\tilde{\mv{V}}_k^H$. By
letting $\mv{A}=\mv{Q}_k$, the proof of Lemma \ref{lemma:1} thus
follows.

\section{Proof of Lemma \ref{lemma:2}}\label{appendix:proof 2}

We prove Lemma \ref{lemma:2} by contradiction. Suppose that there
exist a number of strictly positive $\mu_a$'s such that $A_{\mu}<
\lceil\frac{M-N(K-1)}{M_B}\rceil$. Then, it follows that
$A_{\mu}<\frac{M-N(K-1)}{M_B}$. Since $L=N(K-1)$, it thus follows
that $M_BA_{\mu}<(M-L)$. Let $\mathcal{S}$ denote the set consisting
of the indices corresponding to all the non-zero diagonal elements
in $\mv{B}_{\mu}$, i.e., if $\mu_a>0$ for any $a\in\{1,\cdots,A\}$,
then $(a-1)M_B+i\in\mathcal{S}, i=1,\cdots,M_B$. Note that the size
of $\mathcal{S}$ is denoted by $|\mathcal{S}|=M_BA_{\mu}$. Let
$\mv{E}_k(\mathcal{S})$ and $\mv{F}_k(\mathcal{S}^c)$ denote the
matrix consisting of the rows in
$\tilde{\mv{V}}_k\in\mathbb{C}^{M\times (M-L)}$ with the row indices
given by the elements in $\mathcal{S}$ and $\mathcal{S}^c$,
respectively, where $\mathcal{S}^c$ denotes the complement of
$\mathcal{S}$. Note that $|\mathcal{S}|+|\mathcal{S}^c|=M$ and
$|\mathcal{S}^c|>0$ since $M_BA_{\mu}<(M-L)<M$. From
$\mv{E}_k(\mathcal{S})\in\mathbb{C}^{M_BA_{\mu}\times (M-L)}$ and
$M_BA_{\mu}<(M-L)$, it follows that $\mv{E}_k(\mathcal{S})$ is not
full row-rank. Thus, we could find a vector
$\mv{q}_k\in\mathbb{C}^{(M-L)\times 1}$ with $\|\mv{q}_k\|=1$ such
that $\mv{E}_k(\mathcal{S})\mv{q}_k=\mv{0}$ and
$\mv{F}_k(\mathcal{S}^c)\mv{q}_k\neq\mv{0}$. Accordingly, we have
$\mv{B}_{\mu}\tilde{\mv{V}}_k\mv{q}_k=\mv{0}$ and
$\tilde{\mv{V}}_k\mv{q}_k\neq\mv{0}$. Denote
$\mv{w}_k=\tilde{\mv{V}}_k\mv{q}_k$. Note that the indices of the
non-zero elements in $\mv{w}_k$ belong to $\mathcal{S}^c$. Suppose
that the solution of (P3) is taken as
$\mv{Q}_k^{\star}=p(\mv{q}_k\mv{q}_k^H)$ with $p\geq0$. Substituting
this solution into the objective function of (P3) yields
\begin{align}
&~w_k\log\left|\mv{I}+\mv{H}_k\tilde{\mv{V}}_k\mv{Q}_k^{\star}\tilde{\mv{V}}_k^H\mv{H}_k^H\right|-
{\rm Tr}\left({\mv
B}_{\mu}\tilde{\mv{V}}_k\mv{Q}_k^{\star}\tilde{\mv{V}}_k^H\right) \\
=&~w_k\log\left|\mv{I}+p\mv{H}_k\mv{w}_k\mv{w}_k^H\mv{H}_k^H\right|.
\label{eq:1}
\end{align}
Let $\mv{R}_k=\mv{H}_k\mv{w}_k$. Then, (\ref{eq:1}) can be further
expressed as $w_k\log\left|\mv{I}+p\mv{R}_k\mv{R}_k^H\right|$, whose
value becomes unbounded as $p\rightarrow \infty$ provided that
$\mv{R}_k\mv{R}_k^H\neq\mv{0}$ (which holds with probability one due
to independent channel realizations). Therefore, we conclude that
the presumption that $A_{\mu}< \lceil\frac{M-N(K-1)}{M_B}\rceil$
cannot be true. Lemma \ref{lemma:2} thus follows.

\section{Proof of Lemma \ref{lemma:3}}\label{appendix:proof 3}

We prove Lemma \ref{lemma:3} by contradiction. Suppose that $A^*\geq
(NK+1)$. Let $\mathcal{B}$ be a subset of $\{1,\cdots,A\}$
consisting of the indices of the BSs for which the transmit power
constraints are tight with the optimal solution for (P6). Note that
$|\mathcal{B}|=A^*$. Let $\bar{\lambda}_{k,i}^{\star}$ denote the
optimal solution for (P6), $k=1,\cdots,K$ and $i=1,\cdots,N$. Thus,
we have the following equalities from (P6)
\begin{align}
\sum_{k=1}^K\sum_{i=1}^N\|\mv{v}_k^{\perp}[a,i]\|^2
\bar{\lambda}_{k,i}^{\star}=P,  ~\forall a\in\mathcal{B}.
\end{align}
Accordingly, $\bar{\lambda}_{k,i}^{\star}$'s are the solutions for a
set of $A^*$ linear independent (which holds with probability one
due to independent channel realizations) equations. However, since
$A^*\geq (NK+1)$, we see that the number of equations exceeds that
of unknowns, which is equal to $NK$. Thus, given $P>0$, there exist
no feasible solutions for $\bar{\lambda}_{k,i}^{\star}$'s. We thus
conclude that the presumption that $A^*\geq (NK+1)$ cannot be true.
Lemma \ref{lemma:3} thus follows.

\newpage

\end{document}